\documentstyle[epsf]{aipproc}

\begin{document}
\title{Testing cosmological variability of fundamental constants}

\author{D.~A.~Varshalovich, A.~Y.~Potekhin, and A.~V.~Ivanchik}
\address{Ioffe Physico-Technical Institute, 
194021 St.\,Petersburg, Russia}

\maketitle

\begin{abstract}
One of the topical problems of contemporary physics is a possible
variability of the fundamental constants. 
Here we consider possible variability of two dimensionless constants
which are most important for calculation of atomic and molecular spectra
(in particular, the X-ray ones):
the fine-structure constant $\alpha=e^2/\hbar c$ and
the proton-to-electron mass ratio $\mu=m_p/m_e$.
Values of the physical constants in the early epochs are estimated
directly from observations of quasars --
the most powerful sources of radiation,
whose spectra were formed when the Universe was 
several times younger than now.
A critical analysis of the available results
leads to the conclusion that present-day data do not reveal
any statistically significant evidence for 
variations of the fundamental constants under study.
The most reliable upper limits to possible variation rates
at the 95\% confidence level,
obtained in our work, read:
$$
  |\dot\alpha/\alpha| < 1.4\times 10^{-14}{\rm~yr}^{-1},
\quad
  |\dot\mu/\mu| < 1.5\times10^{-14}{\rm~yr}^{-1}
$$
on the average over the last $10^{10}$ yr.
\end{abstract}

\section*{Introduction}
Contemporary theories (SUSY GUT, superstring and others) 
not only predict the dependence of fundamental physical 
constants on energy\footnote{
The prediction of the theory that 
the fundamental constants depend on the energy of interaction
has been confirmed in experiment.
In this paper, we consider only the space-time variability
of their low-energy limits.}, but also 
have cosmological solutions in which low-energy values of 
these constants vary with the cosmological time. 
The predicted variation at the present epoch is small but non-zero, 
and it depends on theoretical model. 
In particular, 
Damour and Polyakov \cite{Damour} have developed
a modern version of the string theory, whose parameters 
could be determined from 
cosmological variations of the coupling constants and
hadron-to-electron mass ratios. 
Clearly, a discovery of these variations would be
a great step in our understanding of Nature.
Even a reliable upper bound on a possible variation rate
of a fundamental constant presents a valuable tool
for selecting viable theoretical models.

Historically, a hypothesis that the fundamental constants 
may depend on the {\it cosmological time\/} $t$
(that is the age of the Universe) was first discussed 
by Milne \cite{Milne} and Dirac \cite{Dirac}.
The latter author proposed his famous ``large-number hypothesis''
and suggested that the gravitational constant was directly
proportional to $t$. 
Later the variability of fundamental constants
was analyzed, using different arguments, by 
Gamow \cite{Gamow}, Dyson \cite{Dyson}, and others.
The interest in the problem has been revived due to recent
major achievements in GUT and Superstring models (e.g.,~\cite{Damour}).

Presently, the fundamental constants are being measured with a relative
error of $\sim 10^{-8}$. 
These measurements obviously rule out considerable variations of the constants
on a short time scale, 
but do not exclude their changes over the lifetime
of the Universe, $ \sim 1.5\times10^{10}$ years.
Moreover, one cannot rule out the possibility that the constants differ
in widely separated regions of the Universe; this could
be disproved only by astrophysical observations
and different kinds of experiments.

\begin{figure}[!ht]
\begin{center}
\epsfbox[50 470 550 700]{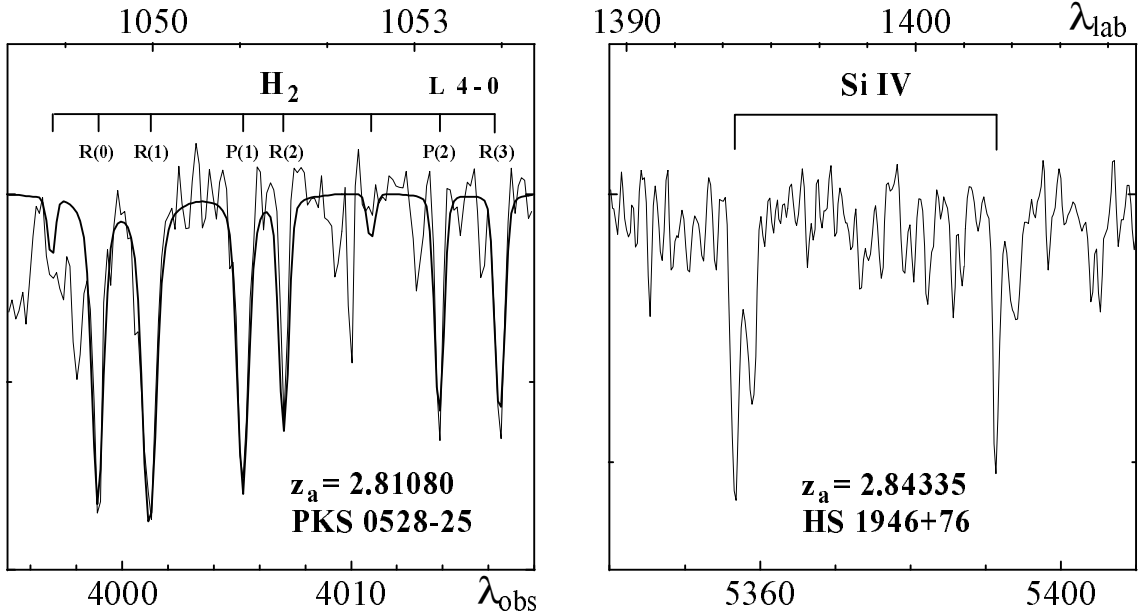}
\end{center}
\caption{Portions of quasar spectra which show
the absorption lines of H$_2$ and Si{\sc\,iv} with large redshifts, 
$z=(\lambda_{\rm obs}-\lambda_{\rm lab})/\lambda_{\rm lab}\approx 2.8$.
The lower horizontal axis gives the wavelengths
in the observer's frame ($\lambda_{\rm obs}$) and the upper
axis gives the wavelengths
in the quasar's frame $\lambda_{\rm lab}$ (in {\AA}).
(a)~The spectrum (thin line) of the quasar PKS 0528--250,
obtained with the 4-meter CTIO telescope (Chile), containing
H$_2$ lines which belong to the L 4--0 branch of the spectrum;
thick line plots the spectral fit.
(b)~The spectrum of the quasar HS 1946+76,
obtained with the 6-meter SAO telescope (Russia), containing
Si{\sc\,iv} doublet lines which correspond to the
$^2$S$_{1/2}\to ^2$P$_{3/2}$ and $^2$S$_{1/2}\to ^2$P$_{1/2}$
transitions.
\label{fig1}
}
\end{figure}

Laboratory experiments cannot trace
possible variation of a fundamental constant
during the entire history of the Universe.
Fortunately, Nature has provided us with a tool 
for direct measuring the physical constants in the early epochs.
This tool is based on observations of quasars,
the most powerful sources of radiation.
Many quasars belong to most distant objects we can observe.
Light from the distant quasars travels to us about $10^{10}$~years. 
This means that the quasar
spectra registered now were formed $\sim 10^{10}$ years ago.
The wavelengths of the lines observed in these spectra
($\lambda_{\rm obs}$) increase compared to their 
laboratory values ($\lambda_{\rm lab}$) in proportion 
$\lambda_{\rm obs}$ = $\lambda_{\rm lab} (1 + z)$, where 
the {\it cosmological redshift\/} $z$ can be used to determine the age 
of the Universe at the line-formation epoch.
In some cases, the redshift is as high as $z\sim 3{-}5$,
so that the intrinsically far-ultraviolet lines
are registered in the visible range.
The examples are demonstrated in Fig.~\ref{fig1}.
Analysing these spectra we may study
the epoch when the Universe was several times younger than now.

Here we review briefly the studies of the space-time
variability of the fine-structure constant $\alpha$ 
and the proton-to-electron mass ratio $\mu$. 

\section*{Fine-structure constant}
Various tests of the fundamental constant variability differ in 
space-time regions of the Universe which they cover. 
{\em Local tests\/} relate to the values of constants 
on the Earth and in the Solar system.
In particular, {\em laboratory tests\/}  
infer the possible variation of certain combinations 
of constants ``here and now'' from comparison of different 
frequency standards. {\em Geophysical tests\/}
 impose constraints 
on combinations of fundamental constants over the past history 
of the Solar system, although most of these constraints are very indirect.
In contrast, {\em astrophysical tests\/} allows one to ``measure'' 
the values of fundamental constants in distant areas of the early 
Universe.

\subsection*{Local tests}

\subsubsection*{Laboratory experiments}

There were a number of laboratory experimets aimed at 
detection of trends of the fundamental constants
with time by comparison of frequency standards which have
different dependences on the constants.
We mention only two of the published experiments.

Comparison of H-masers with Cs-clocks during 427 days
revealed a relative (H--Cs) frequency drift with a rate 
$1.5\times10^{-16}$ per day, while the rates of (H-H) and (Cs--Cs)
drifts (i.e., the drifts between identical standards,
used to control their stability) were less than $1\times10^{-16}$ per day
\cite{Demidov}.
A similar result was found in comparison of a Hg$^+$-clock with 
a H-maser during 140 days \cite{Prestage}:
the rate of the relative frequency drift was less than $(2\pm1)\times10^{-16}$ 
per day.

Such a drift is treated as a consequence of a difference
in the long-term stability of different atomic clocks.
In principle, however, it may be caused by variation of $\alpha$.
That is why it gives an upper limit to the $\alpha$ variation \cite{Prestage}:
$|\dot\alpha/\alpha| \leq 3.7\times10^{-14}{\rm~yr}^{-1}$.

\subsubsection*{Geophysical tests}

The strongest bound on the possible time-variation rate of $\alpha$
was derived in 1976 by Shlyakhter \cite{Shlyakhter}, 
and recently, from a more detailed analysis,
by Damour and Dyson \cite{DD}, who obtained
$|\dot\alpha/\alpha| < 0.7\times 10^{-16}{\rm~yr}^{-1}$,
The analysis was based on measurements of isotope ratios
in the Oklo site in Gabon, where a unique natural uranium 
nuclear fission reactor
had operated 1.8 billion years ago.
The isotope ratios of samarium 
produced in this reactor by the neutron capture reaction 
$^{149}$Sm$+n\to^{150}$Sm$+\gamma$ 
would be completely different, if the energy of the nuclear 
resonance responsible for this capture were shifted at least by 0.1 eV.

Another strong bound, 
$|\dot\alpha/\alpha| < 5\times10^{-15}{\rm~yr}^{-1}$,
was obtained by Dyson \cite{Dyson} 
from an isotopic analysis of natural radioactive decay products 
in meteorites.

A weak point of these tests is their dependence
on the model of the phenomenon,
fairly complex, involving many physical effects.
For instance, Damour and Dyson \cite{DD} estimated possible shift
of the above-mentioned resonance due to the $\alpha$ variation,
assuming that the Coulomb energy of the 
excited  state of $^{150}$Sm$^*$,
responsible for the resonance,
is not less than the Coulomb energy of the
{\em ground\/} state of $^{150}$Sm.
In absence of experimental data on the nuclear state in question,
this assumption is not justified, 
since heavy excited
nuclei often have Coulomb energies smaller than those for
their ground states \cite{Kalvius}.
Furthermore,
a correlation between the constants of strong and electroweak
interactions (which is likely in the frame of modern theory)
might lead to further softening 
of the mentioned bounds by 100-fold, to 
$|\dot\alpha/\alpha| < 5\times10^{-15}{\rm~yr}^{-1}$,
as noted by Sisterna and Vucetich 
\cite{Sisterna}.

In addition, the local tests 
cannot be extended to distant space regions and to the early
Universe, since the law of possible space-time variation of
$\alpha$ is unknown {\it a priory}.
It is the extragalactic astronomy that allows us to study these
remote regions of spacetime,
in particular the regions which were causally disconnected
at the epoch of formation of the observed absorption spectra.

\subsection*{Astrophysical tests}

To find out whether $\alpha$ changed
over the cosmological time, we have studied
the fine splitting of the doublet lines of Si{\sc\,iv}, C{\sc\,iv}, 
Mg{\sc\,ii} and other ions, 
observed in the spectra of distant quasars.
According to quantum electrodynamics, the relative splitting
of these lines $\delta\lambda/\lambda$
is proportional to $\alpha^2$ (neglecting very small corrections).
Consequently, if $\alpha$
changed with time, then $\delta\lambda/\lambda$ would depend on 
the cosmological redshift $z$. 
This method of measuring $\alpha$ in distant regions
of the Universe had been first suggested by Savedoff \cite{Savedoff}
and was used later by other authors.
For instance, Wolfe {\it et al.\/} \cite{Wolfe} derived an estimate
$|\dot{\alpha}/\alpha| < 4\times 10^{-12}\ {\rm yr}^{-1}$ 
from an observation of the Mg{\sc\,ii} absorption doublet at $z = 0.524$.

An approximate formula which relates a deviation
of $\alpha$ at redshift $z$ from its current value, $\Delta\alpha_z$,
with measured $\delta\lambda/\lambda$ in the extragalactic spectra
and in laboratory reads
\begin{equation}
   \Delta\alpha_z \approx {c_r\over2} 
         \left[ {(\delta\lambda/\lambda)_z \over (\delta\lambda/\lambda)_0}
                 -1 \right],
\label{Dalpha}
\end{equation}
where $c_r\sim 1$ takes into account radiation corrections \cite{Dzuba}:
for instance, for Si{\sc\,iv} $c_r\approx0.9$.
  
Many high-quality quasar spectra measured in the last decade
have enabled us to significantly increase the accuracy
of determination of $\delta\lambda/\lambda$ at large $z$.
An example of the spectra observed is shown in Fig.~\ref{fig1}.
For the present report, we have 
selected 
 the results of high-resolution observations \cite{Petitjean,VPI,Outram},
most suitable for an analysis of $\alpha$ variation.
The values of $\Delta\alpha/\alpha$ calculated from these data
according to Eq.~(\ref{Dalpha}) are given in Table~\ref{tab1}.

\begin{table}
\caption{Variation of $\alpha$ value estimated from redshifted
Si{\sc\,iv} fine-splitting doublets.}
\label{tab1}
\begin{tabular}{lddc}
{\bf Quasar}       & $z$      &$\Delta\alpha/\alpha$ & {\bf Ref.} \\
\tableline
HS 1946+76 & 3.050079 &  1.58  & \cite{VPI}  \\
HS 1946+76 & 3.049312 &  0.34  & \cite{VPI}  \\
HS 1946+76 & 2.843357 &  0.59  & \cite{VPI}  \\
S4 0636+76 & 2.904528 &  1.37  & \cite{VPI}  \\
S5 0014+81 & 2.801356 & -1.80  & \cite{VPI}  \\
S5 0014+81 & 2.800840 & -1.70  & \cite{VPI}  \\
S5 0014+81 & 2.800030 &  1.11  & \cite{VPI}  \\
             &          &        &    \\
PKS 0424$-$13 & 2.100027 & -4.51  & \cite{Petitjean}  \\
Q   0450$-$13 & 2.230199 & -1.48  & \cite{Petitjean}  \\
Q   0450$-$13 & 2.104986 &  0.02  & \cite{Petitjean}  \\
Q   0450$-$13 & 2.066646 &  1.03  & \cite{Petitjean}  \\
              &          &        &    \\
J   2233$-$60 & 1.867484 & -1.92  & \cite{Outram}  \\
J   2233$-$60 & 1.869756 & -2.21  & \cite{Outram}  \\
J   2233$-$60 & 1.871074 & -1.41  & \cite{Outram}  \\
J   2233$-$60 & 1.925971 &  1.11  & \cite{Outram}  \\
J   2233$-$60 & 1.941979 &  0.48  & \cite{Outram}  \\
\end{tabular}
\end{table}

As a result, we obtain a new estimate of the possible 
deviation of the fine-structure constant at $z = 2$--4 
from its present ($z = 0$) value:
\begin{equation}
\Delta\alpha/\alpha = 
      (-4.6\pm4.3\,[{\rm stat}]\pm1.4\,[{\rm syst}])\times10^{-5},
\end{equation}
where the statistical error is obtained from the scatter of 
astronomical data (at large $z$) and the systematic one
is estimated
from the uncertainty of the fine splitting measurement
in the laboratory 
\cite{Morton,Kelly}
(at $z=0$, which serves as the reference
point for the estimation of $\Delta\alpha$).
The corresponding upper limit
of the $\alpha$ variation rate averaged over $\sim10^{10}$~yr
is 
\begin{equation}
|\dot{\alpha}/\alpha| < 1.4\times10^{-14}{\rm~yr}^{-1}
\label{alpha-bound}
\end{equation}
(at the 95\% confidence level).
This constraint is much more stringent than those 
obtained from all but one previous astronomical observations. 
The notable exception is presented 
by Webb et al.\ \cite{Webb}, who have analysed 
spectroscopic data of similar quality, but
estimated $\alpha$ from comparison of Fe{\sc\,ii}
and Mg{\sc\,ii} fine-splitted walelengths
in extragalactic spectra and in the laboratory.
Their result indicates
a tentative time-variation of $\alpha$:
$\Delta\alpha/\alpha=(-1.9\pm0.5)\times10^{-5}$ at $z=1.0$--1.6.
Note, however, 
two important sources of a possible systematic error
which could mimic the effect:
(a) Fe{\sc,ii} and Mg{\sc\,ii} lines used
are situated in different orders of the echelle-spectra,
so relative shifts in calibration of the different orders
can simulate the effect of $\alpha$-variation,
and (b) were the relative abundances of Mg isotopes
changing during the cosmological evolution,
the Mg{\sc\,ii} lines would be subjected to an additional
$z$-dependent shift relative to the Fe{\sc\,ii} lines,
quite sufficient to simulate the variation of $\alpha$
(this shift can be easily estimated from recent
laboratory measurements \cite{Pickering}).
In contrast, the method based on the fine splitting
of a line of the same ion species (Si{\sc\,iv} in the above example)
is not affected by these two uncertainty sources.
Thus we believe that the restriction (\ref{alpha-bound})
is the most reliable at present for the long-term history of the Universe.

According to our analysis, 
some theoretical
models are inconsistent with observations.
For example, power laws $\alpha\propto t^n$
 with $n = 1$, $-1/4$, and $-4/3$, 
published by various authors in 1980s, are excluded.
Moreover, the Teller--Dyson's hypothesis on the
logarithmic dependence of $\alpha$ on $t$
\cite{Teller,Dyson} has also been shown to be inconsistent with observations.

Many regions of formation of the spectral lines,
observed at large redshifts in different
directions in the sky, had been causally disconnected at the
epochs of line formation.
Thus, no information could have been exchanged
between these regions of the Universe
and, in principle, the fundamental constants could
be different there.
However, a separate analysis \cite{SpSciRev}
has shown that $\alpha$ value is the same in different directions
in the sky
within the $3\sigma$ relative error 
$|\Delta\alpha/\alpha| < 3\times10^{-4}$.

\section*{Proton-to-electron mass ratio}
The dimensionless constant $\mu = m_{\rm p}/m_{\rm e}$ approximately equals 
the ratio of the constants of strong interaction $g^2/(\hbar c)\sim 14$ 
and electromagnetic interaction $\alpha\approx 1/137.036$, where
$g$ is the effective coupling constant
calculated from the amplitude of nucleon--$\pi$-meson scattering
at low energy.

In order to check the cosmological variability of $\mu$
we have used high-redshift absorption lines of molecular hydrogen H$_2$
in the spectrum of the quasar PKS 0528--250.
This is the first (and, in a sense, unique)
high-redshift system of H$_2$ absorption lines
discovered in 1985 \cite{LV85}.
A study of these objects yields information of paramount importance
on the physical conditions $\sim10^{10}$ years ago.

A possibility 
of distinguishing between 
the cosmological redshift of spectral wavelengths 
and shifts due to a variation of $\mu$ arises from the 
fact that the electronic, vibrational, and rotational energies of 
H$_2$ each undergo a different dependence on the reduced mass of 
the molecule.  Hence 
comparing ratios of wavelengths $\lambda_i$ of various H$_2$
electron-vibration-rotational lines in a quasar spectrum at some 
redshift $z$ and in laboratory (at $z=0$), we can trace
variation of $\mu$. 
The method had been used previously by Foltz {\it et al.} \cite{FCB},
whose analysis was corrected later in our papers \cite{VL93,SpSciRev,VP96}.
In the latter papers,
we calculated the sensitivity coefficients $K_i$
of the wavelengths $\lambda_i$ with respect
to possible variation of $\mu$ and applied a linear regression analysis to 
the measured redshifts of individual lines $z_i$ as function of $K_i$.
An illustration of the wavelength dependences on the mass of 
the nucleus is given in Table \ref{tab2},
where a few resonance wavelengths of hydrogen, deuterium, and
tritium molecules
are listed. One can see that, as the nuclear mass increases,
different wavelengths shift in different directions.
More complete tables, as well as two algorithms 
of $K_i$ calculation, are given in Refs.\ \cite{SpSciRev,Lanzetta}.

\begin{table}
\caption{Comparison of wave\-lengths of electron-vibro-rotational lines
for H$_2$, D$_2$, and $T_2$.}
\label{tab2}
\begin{tabular}{cdddc}
$i$       & $\lambda_i({\rm H}_2)$ & $\lambda_i({\rm D}_2)$ & 
$\lambda_i({\rm T}_2)$ & $K_i$ \\
\tableline
L 0--0 R(1) & 1108.633 & 1103.351 & 1101.021 & $-8.18\times10^{-3}$ \\
L 0--2 R(1) & 1077.697 & 1081.153 & 1082.760 & $+5.35\times10^{-3}$ \\
L 0--9 R(1) &  992.013 & 1015.610 & 1027.218 & $+3.80\times10^{-2}$ \\
\end{tabular}
\end{table}

Thus, if the proton mass in the epoch of line formation
were different from the present value, the
measured $z_i$ and $K_i$ values would correlate:
\begin{equation}
{z_i\over z_k} = 
{(\lambda_i/\lambda_k)_z \over (\lambda_i/\lambda_k)_0} 
\simeq 
1+(K_i-K_k)\left({\Delta\mu\over\mu}\right).
\label{1}
\end{equation}

\begin{figure}[!ht]
\begin{center}
\epsfbox[90 260 610 580]{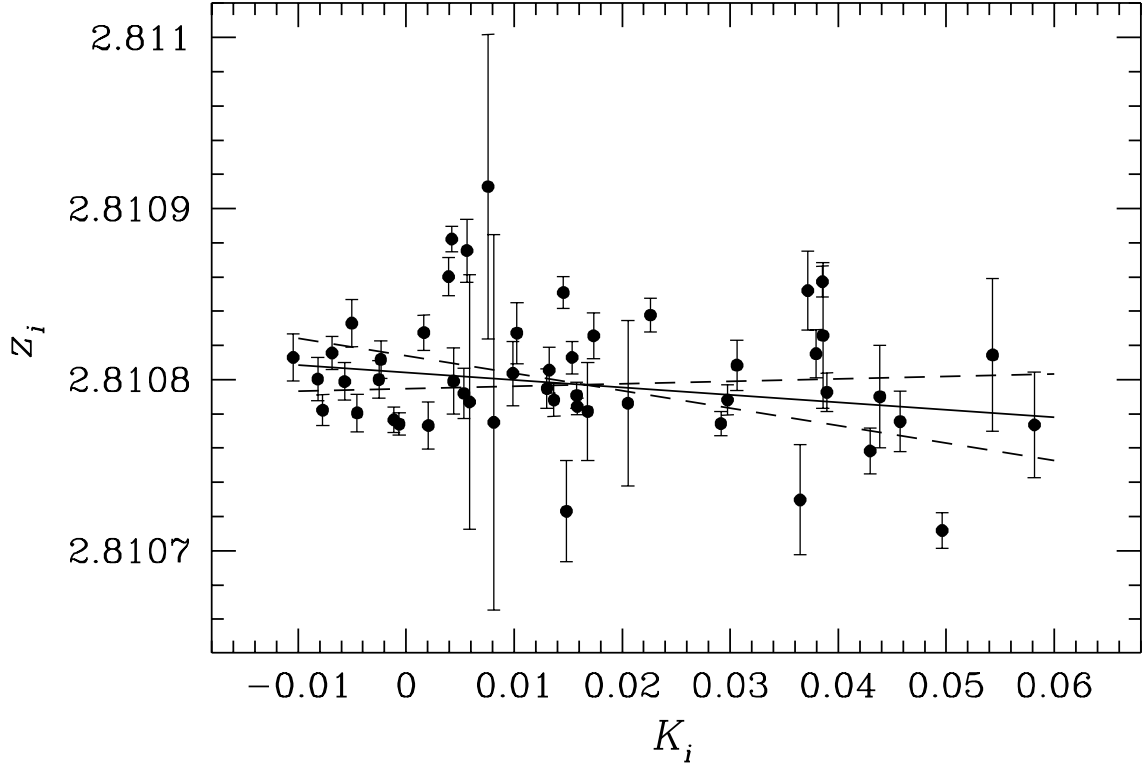}
\end{center}
\caption{Redshift values
inferred from an analysis of separate spectral features
in an H$_2$ absorption system in the spectrum of 
the quasar PKS 0528$-$250, 
plotted vs.\ $\lambda_i(\mu)$-sensitivity coefficients $K_i$.
The slanted solid line shows the most probable regression
and the dashed ones corespond to $\pm1\sigma$ 
deviations of the slope.}
\label{fig2}
\end{figure}

We have performed a $z$-to-$K$ regression analysis using
a modern high-resolution spectrum
of PKS 0528$-$250 \cite{Lanzetta}.
Several tens of the H$_2$ lines have been identified;
a portion of the spectrum which reveales some of the lines
is shown in Fig.~\ref{fig1}.
The redshift estimates for individual absorption lines
with their individual errorbars are plotted in Fig.~\ref{fig2}
against their sensitivity coefficients.
The resulting parameter estimate and $1\sigma$ uncertainty is
\begin{equation}
\Delta \mu/\mu = (-11.5\pm7.6{\rm\,[stat]}\pm1.9{\rm\,[syst]})
      \times 10^{-5}. 
\label{dmu}
\end{equation}
The $2 \sigma$ confidence interval to $\Delta \mu / \mu$ is 
\begin{equation}
 |\Delta \mu / \mu| < 2.0\times 10^{-4}. 
\label{limits}
\end{equation}
Assuming that the age of the Universe is $\sim 15$ Gyr 
the redshift of the H$_2$ absorption system
$z=2.81080$ corresponds to the elapsed time of 13 Gyr 
(in the standard cosmological model). 
Therefore we arrive at the restriction 
\begin{equation}
|\dot{\mu}/\mu|< 1.5\times 10^{-14}{\rm ~yr}^{-1}
\label{constraint}
\end{equation}
on the variation rate of $\mu$, averaged over 90\% of 
the lifetime of the Universe.

\section*{Conclusions}
Despite the theoretical prediction of the time-dependences
of fundamental constants, 
a statistically significant variation of any of the constants
have not been reliably detected up to date,
according to our point of view substantiated above.
The upper limits obtained indicate that the constants
of electroweak and strong interactions
did not significantly change over the last 90\% of the history
of the Universe.
This shows that more precise measurements and observations
and their accurate statistical analyses are required
in order to detect the expected variations of the 
fundamental constants.

{\bf Acknowledgements.}
This work was performed in frames of the Project 1.25
of the Russian State Program ``Fundamental Metrology''
and supported by the grant RFBR 99-02-18232.

\newcommand{\article}[5]{#1, {\it #3} {\bf #4}, #5 (#2).}


\begin{references}
\bibitem{Damour}
\article{Damour, T., and Polyakov, A.M.}{1994}{Nucl.\ Phys.}{B 423}{532}
\bibitem{Milne}
\article{Milne, E.}{1937}{Proc.\ R.\ Soc.}{A 158}{324}
\bibitem{Dirac}
\article{Dirac, P.A.M.}{1937}{Nature}{139}{323}
\bibitem{Gamow}
\article{Gamow, G.}{1967}{Phys.\ Rev.\ Lett.}{19}{759}
\bibitem{Dyson}
\article{Dyson, F.J.}{Cambridge Univ., Cambridge, 1972}{{\rm in} 
Aspects of Quantum Theory\/,
{\rm edited by A.~Salam and E.~P. Wigner,}}{}{p.~213}
\bibitem{Demidov}
\article{Demidov, N.A., Ezhov, E.M., Sakharov, B.A., et al.}{1992}{%
{\rm in} Proc.\ of 6th European Frequency and Time Forum}{}{p.~409}
\bibitem{Prestage}
\article{Prestage, J.D., Tjoelker, R.L., and Maleki, L.}{1995}{%
Phys.\ Rev.\ Lett.}{14}{3511}
\bibitem{Shlyakhter}
\article{Shlyakhter, A.I.}{1976}{Nature}{25}{340}
\bibitem{DD}
\article{Damour, T., and Dyson, F.J.}{1996}{Nucl.\ Phys.}{B 480}{37}
\bibitem{Kalvius}
 \article{Kalvius, G.M., and Shenoy, G.K.}{1974}{%
Atomic and Nuclear Data Tables}{14}{639}
\bibitem{Sisterna}
\article{Sisterna, P.D., and Vucetich, H.}{1990}{%
Phys. Rev. D}{41}{1034}
\bibitem{Savedoff}
\article{Savedoff, M.P.}{1956}{Nature}{264}{340}
\bibitem{Wolfe}
\article{Wolfe, A.M., Brown, R.L., and Roberts, M.S.}{1976}{%
Phys.\ Rev.\ Lett}{37}{179}
\bibitem{Dzuba}
\article{Dzuba,  V.A., Flambaum, V.V., and Webb, J.K.}{e-print:
physics/9808021}{%
Phys.\ Rev.\ A}{}{submitted}
\bibitem{Petitjean}
\article{Petitjean, P., Rauch, M., and Carswell, R.F.}{1994}{%
Astron.\ Astrohys.}{91}{29}
\bibitem{VPI}
\article{Varshalovich, D.A., Panchuk, V.E., and Ivanchik, A.V.}{1996}{%
Pis'ma Astron.\ Zh. (Engl.\ transl.: Astron.\ Lett.)}{22}{8}
\bibitem{Outram}
\article{Outram, P.J., Boyle, B.J., Carswell, R.F., Hewett, P.C., 
and Williams, R.E.}{1999}{%
Mon.\ Not.\ Roy.\ Astron.\ Soc.}{305}{685}
\bibitem{Morton}
\article{Morton, D.C.}{1991}{Astrophys.\ J.\ Suppl.}{77}{119};
(E) {\bf 81}, 883 (1992)
\bibitem{Kelly}
\article{Kelly, R.L.}{1987}{J.\ Phys.\ Chem.\ Ref.\ Data NBS}{16}{Suppl.\,1}
\bibitem{Webb}
\article{Webb, J.K., Flambaum, V.V., Churchill, C.W., 
 Drinkwater, M.J., and Barrow, J.D.}{1999}{%
Phys.\ Rev.\ Lett.}{82}{884}
\bibitem{Pickering}
\article{Pickering, J.C., Thorne, A.P., and Webb, J.K}{1998}{%
Monthly Not.\ R.\ Astron.\ Soc.}{300}{131}
\bibitem{Teller}
\article{Teller, E.}{1948}{Phys.\ Rev.}{73}{801}
\bibitem{SpSciRev}
\article{Varshalovich, D.A. and Potekhin, A.Y.}{1995}{%
Space Sci.\ Rev.}{74}{259}
\bibitem{LV85}
\article{Levshakov, S.A., and Varshalovich, D.A.}{1985}{%
Monthly Not.\ R.\ Astron.\ Soc.}{212}{517}
\bibitem{FCB}
\article{Foltz, C.B., Chaffee, F.H., and Black, J.H.}{1988}
{Astrophys.\ J.}{324}{267} 
\bibitem{VL93}
\article{Varshalovich, D.A. and Levshakov, S.A.}{1993}{JETP Lett.}
{58}{237}
\bibitem{VP96}
\article{Varshalovich, D.A., and Potekhin, A.Y.}{1996}{%
Pis'ma Astron.\ Zh. (Engl.\ transl.: Astron.\ Lett.)}{22}{3}
\bibitem{Lanzetta}
\article{Potekhin, A.Y., Ivanchik, A.V., Varshalovich, D.A., 
Lanzetta, K.M., et al.}{1998}{%
Astrophys.\ J.}{505}{523}
\end{references}
\end{document}